%Paper: hep-th/9503151
%From: Denis V. Juriev <denis@juriev.msk.ru>
%Date: Thu, 23 Mar 95 07:54:44 +0300

\input amstex
\magnification=1200
\font\cyr=wncyr10
\documentstyle{amsppt}
\NoRunningHeads
\define\tni{\operatorname{int}}
\define\so{\operatorname{\frak s\frak o}}
\define\cth{\operatorname{cth}}
\topmatter
\title
Symmetric designs on Lie algebras and interactions of hamiltonian systems
\endtitle
\author
Denis V. Juriev
\endauthor
\affil
Erwin Schr\"odinger International Institute for Mathematical Physics,
Pasteurgasse 6/7, Vienna, A-1090, Austria
\endaffil
\date December 6, 1994\enddate
\abstract
Nonhamiltonian interaction of hamiltonian systems is considered. Dynamical
equations are constructed by use of symmetric designs on Lie algebras.
The results of analysis of these equations show that some class of
symmetric designs on Lie algebras beyond Jordan ones may be useful for a
description of almost periodic, asymptotically periodic, almost asymptotically
periodic, and possibly, more chaotic systems. However, the behaviour of
systems related to symmetric designs with additional identities is simpler
than for general ones from different points of view. These facts confirm a
general thesis that various algebraic structures beyond Lie algebras may be
regarded as certain characteristics for a wide class of dynamical systems.
\endabstract
\endtopmatter
\document
Many important classical hamiltonian dynamical systems are connected with Lie
algebras [1,2] or their nonlinear generalizations [3]. An interaction of
hamiltonian systems may be of different kinds. First, it may be hamiltonian,
i.e. defined by a subsidiary term $\Cal H_{\tni}$ in the hamiltonian and,
possibly, by a certain (maybe rather nonlinear) deformation of initial ("free")
Poisson brackets. Second, it may be nonhamiltonian, i.e. with nonconservative
(nonpotential) forces of interaction, however, one may suppose that it is still
nondissipative, i.e. the total energy is conserved. To describe algebraic
structures governing such interactions is an important unsolved problem of
mathematical physics. There exist, at least, two approaches to the problem.
The first approch associates the nonhamiltonian interaction with certain
deformations or generalizations of the initial algebraic structures (Lie
algebras) such as f.e. isotopic pairs [4] or general (nonlinear) {\rm I}--pairs
[5]. The second approach defines such interaction by use of subsidiary
algebraic structures on Lie algebras.
\pagebreak

\definition{Definition 1 \rm [6]} {\it A design}\footnote{\ An English word
"design" is used here as a rough translation of an original Russian term
"{\cyr uzor}". Probably, such translation is not the best because the English
word "design" is reserved for another mathematical object. However, we shall
use it but only as a translation of the Russian "{\cyr uzor}".\newline} on a
Lie algebra $\frak g$ is a $\frak g$--equivariant algebraic structure on it.
\enddefinition
In the paper [6] the Jordan designs on Lie algebras were considered, and the
related nonhamiltonian dynamics was explored. the results confirmed the
supposition that the formalism of Jordan designs may be a useful tool for an
investigation of asymptotically almost periodic systems.

However, some generalizations of Jordan designs may be useful, too. At least,
there are no any self--evident reasons to be restricted by a Jordan structure
\footnote{\ This argument was formulated by Prof.~D.V.~Alekseevski\v\i \ in a
private talk (ESI, 30 Nov. 1994) with the author.\newline}. So it is rather
natural
to try to verify other algebraic structures as possible "governing" objects
for nonhamiltonian interactions, and to specify the class of dynamical systems,
which is convenient to describe by them.

This paper is devoted to a certain class of designs on Lie algebras, which may
be of an interest.

\definition{Definition 2} {\it A symmetric design\/} on a Lie algebra $\frak g$
with the commutator $[\cdot,\cdot]$ is a $\frak g$--equivariant structure of a
triple system on $\frak g$ with a trilinear operation
$\left<\cdot,\cdot,\cdot\right>$ such that
$\left<X,A,Y\right>=\left<Y,A,X\right>$ ($\forall A,X,Y\in\frak g$) and
$[A,\left<X,A,X\right>]+[X,\left<A,X,A\right>]=0$ ($\forall A,X\in\frak g$).
\enddefinition

\remark{Remark 1} If a Lie algebra $\frak g$ admits two symmetric designs with
trilinear operations $\left<\cdot,\cdot,\cdot\right>'$ and $\left<\cdot,\cdot,
\cdot\right>''$ then it admits an infinite family of symmetric designs with
trilinear operations $\lambda\left<\cdot,\cdot,\cdot\right>'+\mu\left<\cdot,
\cdot,\cdot\right>''$. It means that symmetric designs on a fixed Lie algebra
form a linear space.
\endremark

It seems that a linear space of all symmetric designs on a Lie algebra
possesses
a subsidiary hidden algebraic structure, its unravelling is an important
problem,
which is, unfortunately, out of a general line of the present paper.

\remark{Example 1} Let $\frak A$ be an associative algebra with an involution
$*$. The Lie algebra $\frak g=\{X\in\frak A: X^*=-X\}$ is supplied by, at
least, two structures of symmetric designs: $\left<X,Y,\right>'=XYZ+ZYX$,
$\left<X,Y,Z\right>''=\tfrac12(XZY+ZXY+YXZ+YZX)$.
\endremark

\remark{Example 2} A Lie algebra $\frak g$ with an invariant bilinear form
$\left<\cdot,\cdot\right>$ is supplied by, at least, two structures of
symmetric designs: $\left<X,Y,Z\right>'=\left<X,Y\right>Z+\left<Z,Y\right>X$
and $\left<X,Y,Z\right>''=\left<X,Z\right>Y$.
\endremark

\remark{Example 3} An arbitrary Lie algebra $\frak g$ possesses a symmetric
design with the trilinear operation
$\left<X,Y,Z\right>=-([[X,Y],Z]+[X,[Y,Z]])$.
This symmetric design will be called {\it canonical}. Note that $\left<X,X,X
\right>=0$ in a Lie algebra with the canonical symmetric design\footnote{\ \it
Remark 2. \rm Unfortunately, the author is not acknowledged on an implicit
axiomatic definition of the canonical symmetric triple product. Certainly, its
construction does not claim a presence of a Lie algebra, one may use only a
structure of a Lie triple system (intimately related to abstract symmetric
spaces). Canonical symmetric designs systematically appear in descriptions of
dissipative dynamics (see f.e. [7] and refs wherein).\newline}.
\endremark

Note that the trilinear operations $\left<\cdot,\cdot,\cdot\right>'$ of
examples 1 and 2 are Jordan triple products, considered in [6].

\definition{Definition 3} Let $\frak g$ be a Lie algebra with a symmetric
design $\left<\cdot,\cdot,\cdot\right>$. Differential equations
$$\left\{\aligned
\dot A_t=&\{\Cal H,A_t\}+\alpha\left<B_t,A_t,B_t\right>\\
\dot B_t=&\{\Cal H,B_t\}-\alpha\left<A_t,B_t,A_t\right>
\endaligned\right.$$
where $(A_t,B_t)\in\frak g\oplus\frak g$, $\{\cdot,\cdot\}$ is a Lie--Poisson
bracket, $\Cal H=\Cal H_1+\Cal H_2+\Cal H_{\tni}$ of interaction, are called
{\it the dynamical equations associated with the symmetric design on the Lie
algebra $\frak g$}.
\enddefinition

These equations generalize a certain partial case of dynamical equations
related to Jordan designs [6].

Note that the dynamical equations associated with the canonical symmetric
design look like coupled Landau--Lifschitz equations [7], however, a certain
principal difference exist --- our dynamical equations are linear by the
internal state and quadratic an by an external field whereas the
Landau--Lifschitz
equations and their analogs are linear by an external field and quadratic by
the internal state.
Namely, such form of dynamical equations provides the conservative law for a
whole system.

If $\frak g$ admits a nondegenerate invariant bilinear form $\left<\cdot,
\cdot\right>$, then it  is reasonable to consider an interaction hamiltonian
in the form $a\left<A,B\right>+b\left<\left<A,B,A\right>,B\right>$. In this
case the dynamical equations will be rewritten in the form:
$$\left\{\aligned
\dot A_t=&\{\Cal H_1,A_t\}+a[B_t,A_t]+\alpha\left<B_t,A_t,B_t\right>+
2b\left<\left<B_t,B_t|A_t,A_t\right>\right>\\
\dot B_t=&\{\Cal H_2,B_t\}-a[B_t,A_t]-\alpha\left<A_t,B_t,A_t\right>-
2b\left<\left<B_t,B_t|A_t,A_t\right>\right>
\endaligned\right.$$
where
$$\aligned
\left<\left<A,B|X,Y\right>\right>=&\tfrac12([A,\left<X,B,Y\right>]+
[B,\left<X,A,Y\right>])=\\
&\qquad-\tfrac12([X,\left<A,Y,B\right>]+[Y,\left<A,X,B\right>]).
\endaligned$$

Let's consider the simplest model example of a Lie algebra $\so(3)$. In this
case the linear space of all symmetric designs is two dimensional, and is
spanned by designs of either exaple 1 or example 2. The hamiltonians $\Cal
H_i$ will be linear hamiltonians of rotators $\Cal
H_1=\left<\Omega_1,A\right>$,
$\Cal H_2=\left<\Omega_2,B\right>$, moreover, $\Omega_1=\Omega_2=\Omega$.
This is just a generalization of a situation described in [6].

A symmetric design on $\so(3)$ will be considered as a deformation of the
Jordan design of [6], i.e. $\left<X,Y,Z\right>_{\varepsilon}=\left<X,Y\right>Z+
\left<Y,Z\right>X+2\varepsilon\left<X,Z\right>Y$. Dynamical equations have
the form (cf.[6]):
$$\left\{\aligned
\dot A=&[\Omega,A]+a[B,A]+2\alpha\left<A,B\right>B+2\alpha\varepsilon
\left<B,B\right>A+2b\left<A,B\right>[B,A]\\
\dot B=&[\Omega, B]-a[B,A]-1\alpha\left<A,B\right>A-2\alpha\varepsilon
\left<A,A\right>B-2b\left<A,B\right>[B,A]
\endaligned\right.$$

It is convenient to put $\tau=\left<A,B\right>$, $\rho=\left<A,A\right>+
\left<B,B\right>$, $\sigma=\left<A,A\right>-\left<B,B\right>$. Then $\dot
\rho=0$ whereas
$$\left\{\aligned
\dot\tau=&-2\alpha(1+\varepsilon)\tau\sigma\\
\dot\sigma=&8\alpha\tau^2+2\alpha\varepsilon(\rho^2-\sigma^2)
\endaligned\right.\tag1$$
Let's express $\tau$ via $\sigma$, $\tau=\tfrac1{2\sqrt{\alpha}}\sqrt{
\tfrac{\dot\sigma}2-\alpha\varepsilon(\rho^2-\sigma^2)}$, and substitute the
resulted expression into the equation for $\dot\tau$. A differential equation
on $\sigma$ is derived, namely,
$$\ddot\sigma+4\alpha(1+2\varepsilon)\sigma\dot\sigma+4\alpha^2\varepsilon
(1+\varepsilon)(\rho^2-\sigma^2)\sigma=0.\tag2$$

First, let's investigate a behaviour of the system of differential equations
(1) qualitatively. Such behaviour essentially depends on the value of
$\varepsilon$. For all values of $\varepsilon$ the system has two boundary
critical points $\tau=0$ and $\sigma=\pm\rho$; if $-1,\varepsilon,0$ it has
also two proper critical points $\sigma=0$ and $\tau=\pm\tfrac{\rho\zeta}2$
($\zeta=\sqrt{-\varepsilon}$, $0<\zeta,1$). If $\varepsilon=-1$ there exists a
critical curve $\rho^2=\sigma^2+4\tau^2$.

The linearized system at the boundary critical point $\tau=0$, $\sigma=\pm\rho$
has the form
$$\left\{\aligned
\dot u=&\mp2\alpha\rho(1+\varepsilon)u\\
\dot v=&\mp 4\alpha\varepsilon\rho v
\endaligned\right.$$
($\tau=u$, $\sigma=\pm\rho\mp v$) whereas this system at the proper critical
point $\sigma=0$, $\tau=\pm\tfrac{\rho\zeta}2$ has the form
$$\left\{\aligned
\dot u=&-\alpha\zeta(1-\zeta^2)\rho v\\
\dot v=&8\alpha\zeta\rho u
\endaligned\right.$$
($\tau=\pm(\tfrac{\rho\zeta}2+u)$, $\sigma=v$).

If proper critical points do not exist (i.e. $\varepsilon<-1$ or
$\varepsilon>0$)
one of boundary critical points is attractive and another is repulsive; such
situation is not interesting. On the contrary, if $-1<\varepsilon<0$ the
boundary points are hyperbolic, whereas proper ones are elliptic. Therefore,
only two cases ($\varepsilon=-1$ and $-1<\varepsilon<0$) will be considered.

\remark{Case 1 \rm ($\varepsilon=-1$)} The symmetric design $\left<\cdot,
\cdot,\cdot\right>_{-1}$ is just canonical one.

The system (2) has the form
$$\left\{\aligned
\dot\tau=&0\\
\dot\sigma=&2\alpha(\sigma^2+4\tau^2-\rho^2)
\endaligned\right.$$

So $\tau=\tau_0$ whereas $\sigma=\varkappa\cth 2\varkappa\alpha t$
($\varkappa=\sqrt{\rho^2-4\tau^2_0}$).

Complete dynamical equations have the form
$$\left\{\aligned
\dot A=&[\Omega,A]+(a+2b\tau_0)[B,A]-\alpha(\rho-\sigma)A+2\alpha\tau_0B\\
\dot B=&[\Omega,B]-(a+2b\tau_0)[B,A]-2\alpha\tau_0A+\alpha(\rho+\sigma)B
\endaligned\right.$$
Put $C=(\rho-\varkappa)A+2\tau_0B$ then
$C\underset{t\to\infty}\to\longrightarrow
0$, and hence, $\dot A\sim[\Omega,A]$, $\dot B\sim[\Omega,B]$, so the dynamics
is asymptotically periodic with period $\Omega$ {\sl independent\/} on the
initial conditions.
\endremark

\remark{Case 2 \rm ($-1<\varepsilon<0$)} Let's put $\zeta=\sin\eta/2$ then the
system (1) is written as
$$\left\{\aligned
\dot\tau=&-2\alpha\cos^2\tfrac{\eta}2\tau\sigma\\
\dot\sigma=&8\alpha\tau^2-2\alpha\sin^2\tfrac{\eta}2(\rho^2-\sigma^2)
\endaligned\right.\tag3$$
whereas the equation (2) has the form
$$\ddot\sigma+2\alpha\cos\eta\,\sigma\dot\sigma-\alpha^2\sin\eta(\rho^2-
\sigma^2)\sigma=0.$$

Note that the symmetric design $\left<\cdot,\cdot,\cdot\right>$ is Jordan iff
$\varepsilon=0$ or $\varepsilon=-1/2$. The least case corresponds to
$\eta=\tfrac{\pi}2$. The system (3) is transformed into
$$\left\{\aligned
\dot\tau=&-\alpha\tau\sigma\\
\dot\sigma=&8\alpha\tau^2-\alpha(\rho^2-\sigma^2)
\endaligned\right.\tag5$$
whereas the equation (4) is rewritten as
$$\ddot\sigma-\alpha^2(\rho^2-\sigma^2)\sigma=0.\tag6$$
It describes an evolution of an anharmonic oscillator, the solutions are
expressed via elliptic functions (and hence, the dynamics (5) is periodic).

However, the detailed analysis of the complete dynamical equations seems to
be beyond purely analytic methods and claims special numerical computations.

The situation for arbitrary $\eta\ne\tfrac{\pi}2$ is similar.
\endremark

In spite of the uncompleteness of the drawn picture the obtained  (partially
qualitative) results confirm an initial supposition of
Prof.~D.V.~Alekseevski\v\i \ that some class of symmetric designs on Lie
algebras beyond Jordan ones may be also useful for a description of almost
periodic, asymptotically periodic, asymptotically almost periodic, and
possibly, more chaotic dynamical systems. Such designs, really, may from
only a bounded domain in the whole linear space of all symmetric designs on
a Lie algebra. Also some special points of this domain, really, may have a
greater importance, if they are related todesigns with additional identities
for the trilinear operation (f.e. the Jordan designs). In such points the
dynamics either possesses additional explicitely integrable components and
asymptotically invariant tori or simplifies its behaviour qualitatively.
This fact confirms the general thesis that various algebraic structures beyond
Lie algebras may be regarded as certain implicit (hidden) internal
characteristics for a wide class of dynamical systems.

The author thanks Erwin Schr\"odinger International Institute for Mathematical
Physics, and Prof.~P.~Michor in particular, for a support, stimulating
atmosphere and a kind hospitality.

\enddocument